\documentclass{an}
\usepackage{graphicx}
\usepackage{times}
\usepackage{fancyhdr}
\usepackage{natbib}
\newcommand{\rs}{R_\mathrm{S}}                            % Schwarzschild radius
\newcommand{\rin}{r_\mathrm{in}}                          % disk inner edge
\newcommand{\quot}[1]{`#1'}                               % "..."
\newcommand{\dd}{\mathrm{d}}                              % differential
\newcommand{\pder}[2]{\frac{\partial #1}{\partial #2}}    % partial derivative
\newcommand{\vect}[1]{\mbox{\protect\boldmath$#1$}}       % poloidal vector
\newcommand{\const}{\mathrm{const}}                       % const
\newcommand{\imp}[1]{\textit{#1}}                         % important text
\newcommand{\ratio}[2]{{\textstyle \frac{\,#1}{\,#2}}}    % rational ratio

\bibpunct{(}{)}{;}{}{}{,}
\begin{document}

\title{A possible mechanism for QPOs modulation\\in neutron star sources}

\author{Ji\v{r}\'{\i} Hor\'ak}
\institute{Astronomical Institute, The Czech Academy of Sciences, 
    Bo\v{c}n\'{\i}~II, CZ-140\,31~Prague, Czech Republic}

\date{Received; accepted; published online}

\abstract{It was pointed out by \citet{pac87} that the X-ray luminosity of
accreting neutron stars is very sensitive to the physical properties of
the accretion flow close to  the innermost stable circular orbit. The
X-ray radiation is dominated by that emitted in the boundary layer, where
accreted matter hits a star surface. The X-ray luminosity of the
boundary layer is proportional to the local accretion rate. In this
note, we estimate local accretion rate variations from the disk that
undergoes non-stationary axisymmetric perturbations. The perturbations 
are given by the poloidal-velocity potential. We obtain a simple formula 
describing the modulation of the accretion rate for the particular case
of  global vertical disk oscillations that have been recently studied by
\citet{abr05}.
}

\correspondence{horak@astro.cas.cz}

\maketitle

%================================================================================
\section{Introduction}
\label{sec:intro}
%================================================================================
The quasi-periodic oscillations (QPOs) appear in the light curves of
more than 20 bright low-mass X ray-binaries (LMXBs) with accreting
neutron stars \citep{klis00}. Much attention is attracted to the
kilohertz QPOs because their frequencies are comparable with orbital
frequencies in the innermost parts of the accretion disks. The orbital
frequency of a particle orbiting the neutron star of mass $M$ at the
innermost stable circular  orbit (ISCO) is
$\nu_\mathrm{ISCO}=1580(1+0.75j)~\mathrm{Hz}\times1.4M_\odot/M$
\citep{klu90}.

Many models have been proposed to explain the excitation mechanism of
QPOs and subsequent modulation of the X-ray signal (see
\citealt{klis00} and \citealt{mcc03} for a detailed  discussion of
observations and models). Recently, it has been suggested that the high
frequency QPOs arise from a resonance between two  oscillation modes of
the innermost part of the accretion disk \citep{klu01,abr01}. In the
case of the neutron-stars sources, the modulation  of the X-ray
radiation may originate in the modulation of the local accretion  rate
\citep{klu04}.

In LMXBs that are not pulsars, the magnetic field of the neutron star 
is sufficiently weak, allowing the accretion disk to extend down to
ISCO.  The strongest X-ray radiation then originates in the boundary
layer, where accreted material hits the star surface. Depending on the
star radius $R_\star$, the amount of energy released in the boundary
layer exceeds that radiated by the whole disk. It gives about 69\% of
the  total luminosity if $R_\star=3\rs$, or even 86\% if
$R_\star=1.5\rs$. 

In this context, \citet{pac87} pointed out that a variability of  X-ray
luminosity of accreting neutron stars may be governed by  physical
properties of the accretion flow close to ISCO. In Einstein gravity, the
inner edge of the pressure supported thick accretion disks is slightly
bellow ISCO \citep{abr85}. The material is accreted from the disk
through a narrow potential nozzle onto the neutron star. Obviously, if
the innermost part of the disk is not stationary but is a subject to
some  oscillations then the fine structure of the flow at the inner disk
edge is significantly changed. This strongly affects  the accretion rate
through the nozzle and the resulting X-ray luminosity of the boundary
layer. This scenario is in agreement with the recent observations of
\citet{gil03} and more recently \citet{rev05} that strongly point to the
fact that neutron-star QPOs are modulated in the boundary layer.  

In sections \ref{sec:ISCO} and \ref{sec:stationary} we briefly summarize equations
important for the disk structure close to ISCO and reproduce the
calculations of the accretion rate through the inner edge of the
stationary disk. Then in section \ref{sec:perturbed} we calculate the
accretion rate from the disk that  is subject to nonstationary
axisymmetric perturbations. We derive a simple  formula for the
accretion rate modulation of a vertically oscillating disk.

%================================================================================
\section{Disk structure close to ISCO}
\label{sec:ISCO}
%================================================================================
We consider an axisymmetric thick disk made of a perfect fluid
surrounding a neutron star of mass $M$. The dynamics of the fluid is
governed by Euler equation, poloidal  component of which takes the form
\begin{equation}
  \pder{\vect{v}}{t}+\vect{v}\cdot\nabla\,\,\vect{v}-
  \frac{\ell^2}{r^2}\vect{e}_r
  +\frac{\nabla p}{\rho}-\nabla\Phi=0,
  \label{eq:Euler}
\end{equation}
where the bold-face letters refer to the poloidal part of the vectors,
$\vect{a}\equiv(a^r,a^z)$, $\Phi$ is a gravitational potential, $r$
denotes radial coordinate (we employ the cylindrical coordinates
$\{r,\phi,z\}$, with the origin coinciding with the center of the star)
and $p$, $\rho$ and $\ell$ are the pressure, density and the angular
momentum of the orbiting flow respectively (in general all depend on
$r$  and $z$). The azimuthal component of the Euler equation gives
conservation of angular momentum,
\begin{equation}
  \pder{\ell}{t} + \vect{v}\cdot\nabla\ell = 0.
  \label{eq:Eulerl}
\end{equation}
We assume that the angular momentum is constant in the whole volume of
the disk, $\ell(r,z)=\ell_0$, and that the fluid obeys the polytropic
equation of state, $P=K\rho^{1+1/n}$, where $K$ and $n$ are polytropic
constant and polytropic  index, respectively. In addition, we assume
that the poloidal velocity  $\vect{v}=(v^r,v^z)$ can be derived from the
potential $\chi$.  Hence, the equation (\ref{eq:Eulerl}) is satisfied
automatically and the equation  (\ref{eq:Euler}) can be further
integrated to Bernoulli equation,
\begin{equation}
  \pder{\chi}{t}+\frac{v^2}{2}+h+\mathcal{U}=\mathrm{const}\equiv
  \mathcal{U}_\mathrm{S}.
  \label{eq:Bernoulli}
\end{equation}
Here we introduced the poloidal-velocity potential by
$\vect{v}=\nabla\chi$,  the enthalpy of the fluid $h\equiv\int_0^p\dd
p/\rho=nK\rho^{1/n}$ and  the effective potential
$\mathcal{U}=\Phi(r,z)+\ell_0^2/2r^2$. 

As a model of a strong gravitational field of the star, we use the
pseudo-Newtonian potential $\Phi(r)=-GM/(R-\rs)$, where
$R\equiv\sqrt{r^2+z^2}$ and $\rs$ is Schwarzschild radius. It was
introduced by \citet{pac80} and allows us to model general relativistic
effects using Newtonian   calculations with remarkable simplicity.
Particularly, it gives a correct position of the marginally stable orbit
at $r=r_\mathrm{ISCO}=3\rs$ and well reproduces  the Keplerian angular
momentum of test particles orbiting the star,
$\ell_\mathrm{K}=\sqrt{GMr^3}/(r-\rs)$. The angular momentum is not a
monotonic function of $r$, as it is in Newtonian gravity
($\rs\rightarrow0$). Instead, it has a minimum at the marginally stable
orbit. 

\begin{figure}
\includegraphics[width=0.5\textwidth]{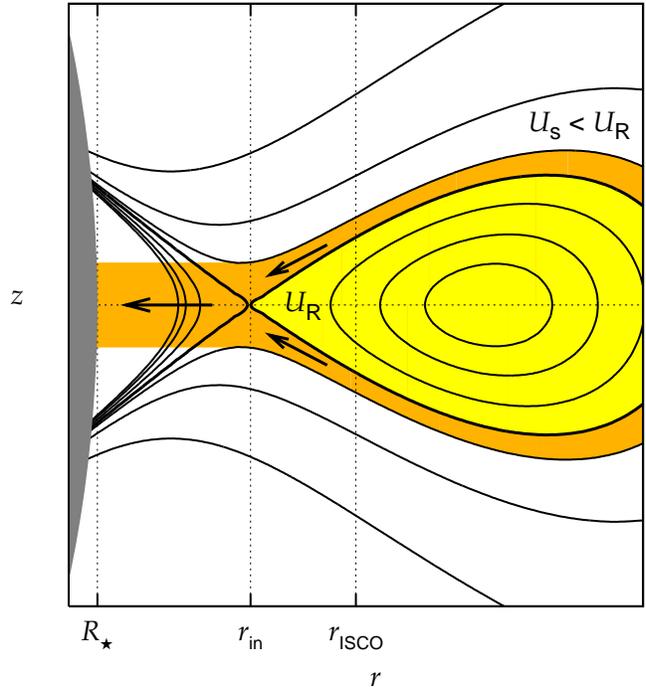}
\caption{Accretion from a thick stationary accretion disk. The position
of the disk inner edge and the shape of the equipotentials are
determined by the distribution  of the fluid angular momentum. Here we
consider the simplest case of the constant distribution,
$\ell(r,z)=\ell_0$. The plot shows projections of the equipotential 
surfaces to the poloidal plane (solid lines) and the distribution of the fluid 
(shaded region).  The matter that
overflows the Roche lobe (the equipotential surface that crosses itself)
is  accreted onto the neutron star.}
\label{fig:accretion}
\end{figure}

The structure of the stationary disk is shown in
Figure~\ref{fig:accretion}. The inner edge is at radius $r=\rin$, where
the angular momentum of the flow equals the Keplerian value,
$\ell_0=\ell(\rin)$. The Lagrange point is at coordinates $[\rin,0]$.
The equipotential surface that crosses  itself at the Lagrange point is
called Roche lobe. The corresponding value of the effective potential is
denoted $\mathcal{U}_\mathrm{R}$ The equilibrium configuration exists  
only if the surface of the torus is inside  the Roche lobe, e.g. when
$\mathcal{U}_\mathrm{s} \leq\mathcal{U}_\mathrm{R}$. \citep{boy65,abr78}

Otherwise, the dynamical equilibrium is impossible and the overflowed
matter will be accreted  through the potential nozzle onto the star.

%================================================================================
\section{Stationary flow}
\label{sec:stationary}
%================================================================================

The stationary accretion rate for Roche overflow was first calculated by
\citet{koz78}, who used Einstein's theory. Here we closely follow the
Newtonian calculations  of \citet{abr85}. We consider a small overflow,
so that all quantities can be expanded to the second order in the
vicinity of the Lagrange point $L$. Particularly, the vertical profile
of the enthalpy can be expressed as
\begin{equation}
  h(\rin,z)\,=\,h^\star-\frac{1}{2}\kappa^2 z^2,\quad\
  \kappa^2\equiv-\left(\frac{\partial^2 h}{\partial z^2}\right)_{\!L},
  \label{eq:h0}
\end{equation}
where $h^\star\equiv h(\rin,0)$ denotes a maximal value of the enthalpy
on the cylinder $r=\rin$. The linear order does not contribute because
the flow is symmetric with respect to the equatorial plane. The
thickness of the inner edge is $H=\sqrt{2h^\star}/\kappa$. Close to
$r=\rin$ the accretion flow becomes transonic.  After \citet{abr85}, we
assume that the radial velocity of the flow equals to the local sound
speed and that the vertical component of the velocity is negligible
compared  to the radial one. This significantly simplifies the solution
because it allows us to  express the poloidal velocity using the
enthalpy,
\begin{equation}
  \vect{v}\,\,=\,\,\sqrt{\frac{h}{n}}\,\,\vect{e}_r.
  \label{eq:velocityanzatz}
\end{equation}

The local mass flux through the nozzle is $\dot{m}=\rho v^r=\rho
c_\mathrm{s}= h^{n+1/2}/K^n(1+n)^n n^{1/2}$ and the integration over the
cylinder $r=\rin$ gives the total mass flux in therms of the central
enthalpy $h^\star$
\begin{eqnarray}
   \dot{M}\!\!\!&=&\!\!\!\int_0^{2\pi}r_1\,\dd\phi\int_{-H}^H\dot{m}\dd z
   \nonumber\\
   \!\!\!&=&\!\!\!(2\pi)^{3/2}\frac{\rin}{n^{1/2}}
   \left[\frac{1}{K(n+1)}\right]^n
   \frac{\Gamma(n+3/2)}{\Gamma(n+2)}\,\frac{(h^\star)^{n+1}}{\kappa},
   \label{eq:Min}
\end{eqnarray}
where $\Gamma(x)$ is the Euler gamma function.

In the Bernoulli equation (\ref{eq:Bernoulli}) we keep the term $v^2/2$ 
and neglect only the time derivative because of stationarity of the
flow. We obtain 
\begin{equation}
  \frac{v^2}{2}+h+\mathcal{U}=\left(1+\frac{1}{2n}\right)h+\mathcal{U}
  \,=\,\mathcal{U}_\mathrm{S}.
  \label{eq:Bernoulli2}
\end{equation}
The parameter $\kappa$ that determines the shape of the enthalpy profile
can be expressed using a derivative of the effective potential. This 
introduces the vertical epicyclic frequency $\omega_z$ to the problem.
From equation (\ref{eq:Bernoulli2}) we obtain
\begin{equation}
   \kappa^2=
   \left(\frac{n}{n+1/2}\right)\omega_z^2,
   \quad\quad
   \omega_z=
   \left(\frac{\partial^2\mathcal{U}}{\partial z^2}\right)_{\!L}\,.
   \label{eq:kappa}
\end{equation}
By substituting the equations (\ref{eq:Bernoulli2}) and (\ref{eq:kappa})
and introducing
$\Delta\mathcal{U}\equiv\mathcal{U}_0-\mathcal{U}_\mathrm{S}$ we finally
recover the  result obtained by \citet{abr85},
\begin{eqnarray}    
  \dot{M}\!\!&=&\!\!A(n)\,\frac{\rin}{\omega_z}\,
  \Delta\mathcal{U}^{n+1}, \\
  A(n)\!\!&\equiv&\!\!(2\pi)^{3/2}\left[\frac{1}{K(n+1)}\right]^n 
  \left[\frac{1}{n+1/2}\right]^{n+1/2}\times
  \nonumber\\
  \!\!&\times&\!\!\frac{\Gamma(n+3/2)}{\Gamma(n+2)}.
\end{eqnarray}   

%================================================================================
\section{A perturbed flow}
\label{sec:perturbed}
%================================================================================

Now, we suppose that the disk is disturbed and oscillates. In that case,
the accretion flow will not  be stationary anymore and in order to
described the flow we  must use the Bernoulli equation
(\ref{eq:Bernoulli}) in the full form. The presence of the
\quot{non-stationary} term $\partial\chi/\partial t$ breaks however the
correspondence between the enthalpy and the effective potential. The
equipotential surfaces and the surfaces of constant enthalpy will not
coincide anymore. If the oscillations are a small perturbation, we can
expand the Bernoulli equation in the vicinity of the stationary flow
considered above.

We suppose that the velocity potential can be expressed as
\begin{equation}
\chi(\vect{r},t)=\chi_{(0)}(\vect{r})+\epsilon\chi_{(1)}(\vect{r},t),
\end{equation}
where the subscript \quot{(0)} refers to the stationary flow and the
dimensionless  parameter $\epsilon$ characterizes strength of the
perturbation. We assume $\epsilon\ll1$. Then, using the definition
$\vect{v}(\vect{r},t)=\nabla\chi(\vect{r},t)$ we find
\begin{eqnarray}
  v^2 &=& v_{(0)}^2+2\epsilon\vect{v}_{(0)}\cdot\vect{v}_{(1)}
  +\epsilon^2 v_{(1)}^2
  \nonumber \\
  &=& c_\mathrm{s}^2+2\epsilon c_\mathrm{s}
  \frac{\partial\chi_{(1)}}{\partial r}+
  \epsilon^2\left[
  \left(\frac{\partial\chi_{(1)}}{\partial r}\right)^{\!2}\!\!\!+\!
  \left(\frac{\partial\chi_{(1)}}{\partial z}\right)^{\!2}
  \right].
\end{eqnarray}
The enthalpy is also affected by the perturbation. The new value can be
approximated by an expansion in the parameter $\epsilon$
\begin{equation}
  h = h_{(0)} + \epsilon h_{(1)} + \epsilon^2 h_{(2)} + 
  \mathcal{O}(\epsilon^3).
\end{equation}
By substituting into the Bernoulli equation (\ref{eq:Bernoulli}) and
equating  coefficients of same powers of $\epsilon$, we get
\begin{eqnarray}
  h_{(1)}&=&-\frac{\partial\chi_{(1)}}{\partial t}-
  \left({\frac{\bar{h}}{n}}\right)^{1/2}
  \frac{\partial\chi_{(1)}}{\partial r},
  \label{eq:h1} \\
  h_{(2)}&=&-\frac{1}{2}\left[
  \left(\frac{\partial\chi_{(1)}}{\partial r}\right)^{\!2}\!\!+
  \left(\frac{\partial\chi_{(1)}}{\partial z}\right)^{\!2}\right].
  \label{eq:h2} 
\end{eqnarray}
This way all thermodynamic quantities are expressed using the
poloidal-velocity potential.

\begin{figure*}
\begin{center}
\resizebox{\hsize}{!}{
\includegraphics[width=0.49\textwidth]{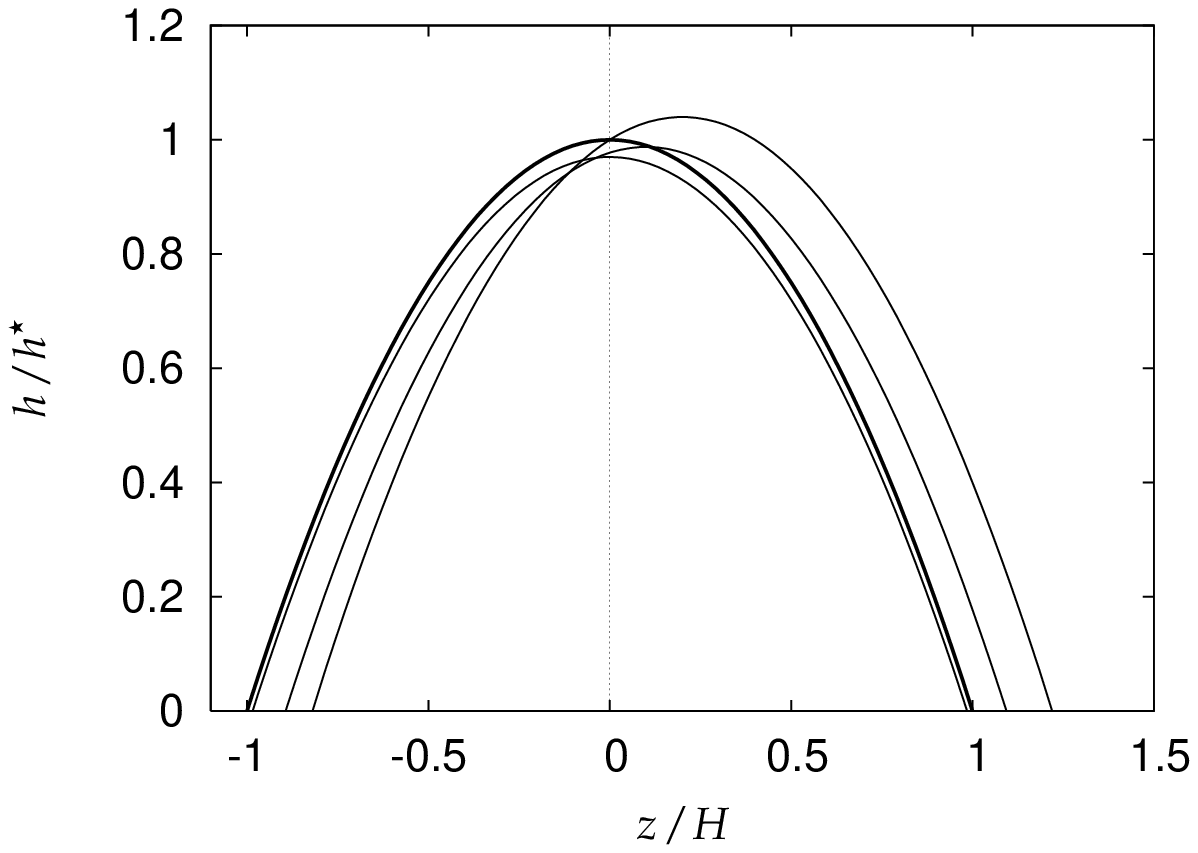}
\hfill
\includegraphics[width=0.49\textwidth]{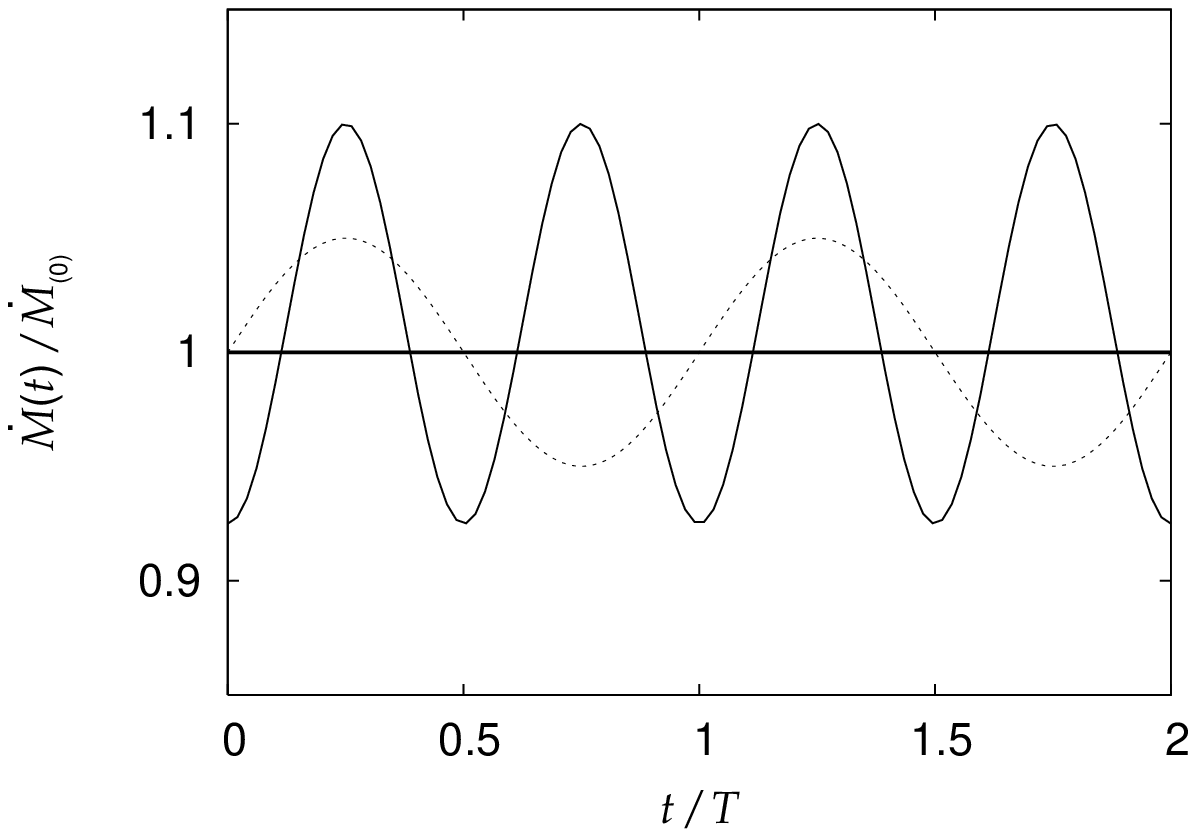}}
\end{center}
\caption{Left: The vertical profiles of the enthalpy $h(\rin,z)$ on the
cylinder $r=\rin$ during vertical disk oscillations (thin lines). The
amplitude of oscillations is $\delta Z=0.2 H$ and we chose the
polytropic  index of the fluid $n=1.5$. The figure captures profiles
with the enthalpy maxima at $\delta z=0,0.1 H$  and $0.2 H$. The
enthalpy profile for the unperturbed stationary disk is also shown
(thick line). Right: The modulated accretion rate from the oscillating
disk (thin solid line). The accretion rate for the stationary disk  is
plotted by thick line. Time is rescaled by the period of oscillations.
For reference we plot also the phase of disk oscillations (dotted line)
The accretion rate is modulated with twice the frequency  of
oscillations.}
\label{fig:profiles}
\end{figure*}

To progress further, we need a particular form of the perturbation
$\chi_{(1)}(\vect{r},t)$. This is a difficult global problem that often
involves numerical calculations. Several authors studied it under
different simplifications. For example, \citet{bla85} gives all possible
modes (e.g. eigenfrequencies and eigenfunctions) of slender-torus
oscillations.  In this limit the size of the  stationary torus is small
enough that the enthalpy can be approximated by a quadratic function in 
the whole torus. \citet{bla85} considered the Newtonian gravitational
field. Recently, \citet{klu02} reconsidered  the problem in 
general relativity potential and pointed to the existence of a particular
mode when the torus moves rigidly up and down across the equatorial
plane (see also \citealp{abr05} for more detailed calculations). 
The eigenfrequency of this mode is equal to the vertical
epicyclic frequency $\omega_z$. The presence of rigid modes in a torus
oscillations has been found also in recent numerical simulations (e.g.
\citealt{lee04}; \citealt{rub05}).

In the following, we model vertical disk oscillations by a simple ansatz
for the poloidal-velocity potential
\begin{equation}
  \chi_{(1)}=z v_z\cos\omega t,
  \label{eq:ansatz}
\end{equation}  
where $\omega$ is the frequency of the oscillations. Calculating
velocity perturbation, we find
\begin{equation}
  \vect{v}_{(1)}=v_z\vect{e}_z\cos\omega t\,.
\end{equation}
Hence, $\epsilon v_z=\const$ can be interpreted as the amplitude of the
vertical velocity. Equations (\ref{eq:h1}) and (\ref{eq:h2}) give
\begin{equation}
  h_{(1)}=z v_z \omega \sin\omega t, \quad
  h_{(2)}=-\frac{1}{2} v_z^2 \cos^2\omega t.
\end{equation}
The vertical profile of the enthalpy at $r=\rin$ reads
\begin{eqnarray}
  h(\rin,z,t)\!\!&=&\!\!h^\star-\kappa^2 z^2 + \epsilon z v_z\omega\sin\omega t-
  \nonumber\\
  \quad
  \!\!&-&\!\!
  \ratio{1}{2}\epsilon^2 v_z^2\cos^2\omega t + 
  \mathcal{O}(\epsilon^3)
\end{eqnarray}
that is quadratic in the variable $z$. The position of the enthalpy
maximum on the cylinder  $r=\rin$ is shifted from $z=0$ to height
$\delta z(t)$ given as
\begin{equation}
  \delta z(t)= \delta Z \sin\omega t, \quad 
  \delta Z = \epsilon\,\frac{\omega v_z}{\kappa^2}.
\end{equation}
We can interpret $\delta Z$ as the amplitude of the oscillations. Also
the value of enthalpy in the maximum differs from the stationary case by
\begin{eqnarray}   
  \delta h^\star\!\!&\equiv&\!\!h(\rin,\delta z)-h^\star 
  \nonumber\\
  \!\!&=&\!\!
  \frac{1}{2}\kappa^2\left[\delta z^2-\frac{\kappa^2}{\omega^2}
  (\delta Z^2-\delta z^2)\right]+\mathcal{O}(\epsilon^3).
  \label{eq:dh}
\end{eqnarray}  

According to equation (\ref{eq:Min}) the actual accretion rate depends
on the maximal enthalpy as  $\dot{M}\propto (h^\star)^{n+1}$. This
relation can be applied also in the case of vertical oscillations
because the $z$-dependence of enthalpy on the cylinder $r=\rin$ can be 
approximated by a quadratic function also in this case and the oscillations  
do not contribute to the radial velocity of accreted matter. Hence, using 
equations (\ref{eq:h0}), (\ref{eq:kappa}) and (\ref{eq:dh}) and assuming that 
the frequency of oscillations equals to the local vertical epicyclic
frequency, $\omega=\omega_z$, we arrive at our final result
\begin{eqnarray}
  \frac{\delta\dot{M}}{\,\,\,\dot{M}_{(0)}}\!\!\!&=&\!\!\!(n+1)
  \frac{\delta h^\star}{h^\star}
  \nonumber\\
  \!\!\!&=&\!\!\!
  \frac{2-p}{2-2p}\left[(1+p)\frac{\delta z^2}{H^2}-p\frac{\delta Z^2}{H^2}\right],
\end{eqnarray}
where $\delta M\equiv\dot{M}-\dot{M}_{(0)}$ and $p\equiv n/(n+1/2)$. 

Figure \ref{fig:profiles} shows the result. The enthalpy profiles
$h(\rin,z)$ are shown for several values of $\delta z$ in the left
panel. The amplitude of oscillations is $\delta Z/H=0.3$. The right
panel shows the modulation of the accretion rate from the oscillating
disk. The time is rescaled by the oscillation period, $T=2\pi/\omega_z$.
Finally, the time-averaged accretion rate is given by
\begin{equation}
  \frac{\langle\delta\dot{M}\rangle}{\,\,\,\dot{M}_{(0)}}=
  \ratio{1}{4}(2-p)\frac{\delta Z^2}{H^2}
\end{equation}
that is positive for reasonable values of $n$.

%================================================================================
\section{Conclusions}
%================================================================================
In this note we studied the accretion rate from a non-stationary
pressure supported accretion disk that undergoes the vertical
axisymmetric oscillations. The oscillations were modelled by a simple
ansatz for the perturbation of poloidal-velocity field. We believe,
however, that  several features would be present also in more
sophisticated (perhaps numerical) solutions: (1) the first correction to
the stationary accretion rate is of the quadratic order in both the
actual perturbation $\delta z$ and the amplitude $\delta Z$. This is
probably because of  the symmetry of the stationary flow with respect to
the equatorial plane. Hence, the frequency  of the modulation must be
twice the oscillation frequency. (2) The accretion rate is maximal when
the disk reaches the maximal amplitude $\delta z=\delta Z$. (3) The
averaged accretion  rate from the periodically perturbed flow is greater
than the that of the stationary flow.

\acknowledgements
The author is grateful to Marek Abramowicz for the invitation to
participate in \imp{Nordita  days on QPOs} conference and also to Wlodek
Klu{\'z}niak and Vladim{\'i}r Karas  for helpful discussion about the
subject. This work was supported  by the GAAV grant IAA 300030510.

%================================================================================

\end{document}